\documentclass[a4paper,twoside]{article}

\usepackage{epsfig}
\usepackage{subcaption}
\usepackage{calc}
\usepackage{amssymb}
\usepackage{amstext}
\usepackage{amsmath}
\usepackage{amsthm}
\usepackage{url}
\usepackage{multicol}
\usepackage{pslatex}
\usepackage{apalike}
\usepackage{algorithm2e}
\usepackage[bottom]{footmisc}
\usepackage{xcolor}
\usepackage{epstopdf}
\usepackage{SCITEPRESS}     

\begin{document}

\title{Multi-face emotion detection for effective Human-Robot Interaction}


\author{\authorname{Mohamed Ala Yahyaoui\sup{1}, Mouaad Oujabour\sup{1}\,  Leila Ben Letaifa \sup{1}\orcidAuthor{0000-0002-0474-3229} and Amine Bohi \sup{2}\orcidAuthor{0000-0002-2435-3017}}
\affiliation{\sup{1}CESI LINEACT Laboratory, UR 7527, Vandoeuvre-lès-Nancy, 54500, France}
\affiliation{\sup{2}CESI LINEACT Laboratory, UR 7527, Dijon, 21800, France}
\email{\{mayahyaoui,moujabour,lbenletaifa, abohi\}@cesi.fr}
}
 
\keywords{Emotion detection, Facial expression recognition, Human-Robot Interaction, Deep learning, Graphical user interface. 
}

\abstract{The integration of dialogue interfaces in mobile devices has become ubiquitous, providing a wide array of services. As technology progresses, humanoid robots designed with human-like features to interact effectively with people are gaining prominence, and the use of advanced human-robot dialogue interfaces is continually expanding. In this context, emotion recognition plays a crucial role in enhancing human-robot interaction by enabling robots to understand human intentions.
This research proposes a facial emotion detection interface integrated into a mobile humanoid robot, capable of displaying real-time emotions from multiple individuals on a user interface. To this end, various deep neural network models for facial expression recognition were developed and evaluated under consistent computer-based conditions, yielding promising results. Afterwards, a trade-off between accuracy and memory footprint was carefully considered to effectively implement this application on a mobile humanoid robot.}

\onecolumn \maketitle \normalsize \setcounter{footnote}{0} \vfill

\section{\uppercase{Introduction}}
\label{sec:introduction}

The rapid advancement of technology in recent years has accelerated research in robotics, with a particular emphasis on humanoid robots. Designed to resemble humans in body, hands, and head, humanoid robots are increasingly capable of sophisticated interactions with people, including recognizing individuals and responding to commands. This human-like form and behavior make them particularly well-suited for applications in human-computer interaction, serving as effective platforms for studying and improving user engagement and interaction dynamics.
Current examples of humanoid robots include Honda's ASIMO \cite{hirose_2007}, known for its advanced mobility and dexterity; Blue Frog Robotics' Buddy \cite{peltier_2017}, designed for social interaction and domestic assistance; and Aldebaran Robotics' NAO \cite{Gouaillier_2009}, recognized for its versatility in research and educational settings. These robots showcase the diversity of roles humanoid robots can play, from companionship and entertainment to education and beyond.

Developing emotional intelligence in robots is relevant as they increasingly participate in social settings. Indeed, beyond performing physical tasks, enhancing robots' ability to perceive, interpret and respond to human needs is essential for effective social Human-Robot Interaction (HRI) and Human-Robot Collaboration (HRC).

In the realm of social robotics, integrating sensors such as microphone for "mouth" or camera for 'eyes' into the humanoid robot, enables the robot to capture human emotions in real-time, and to adapt its response and behavior accordingly \cite{Justo2020,olaso2021,palmero2023exploring}. This capability enhances their utility in various applications and facilitates engagement and intuitive interaction experiences between robots and humans.
Detecting emotions from camera starts with face detection, which involves identifying and locating human faces within images or video frames. This process includes pre-processing images, extracting distinct facial features, classifying regions as faces or non-faces, refining detection accuracy, and handling variations in lighting, occlusions, poses and scales. 
Face emotion recognition (FER) employs computer vision and machine learning techniques to analyze human emotions from face. 

Often, emotion recognition systems deals with only one user while he is communicating with a machine. However, multiple users can communicate simultaneously with it. Multi-face emotion recognition is particularly valuable across various scenarios. For instance, at a comedy club, it provides real-time feedback to comedians, manages lighting and sound, interacts with the audience, and detects disruption.


In this work, we present a complete facial emotion recognition interface and its deployment in a mobile humanoid robot. The proposed interface can display emotions from multiple individuals in real-time within an advanced user interface. To achieve this, several deep neural network models have been developed and evaluated under the same conditions. 
Then a tradeoff between system accuracy and model size have been considered in order to implement the optimal solution into a humanoid robot. The model's performance and its confidence interval also guided this choice of solution.

The remainder of this paper is structured as follows. Section \ref{sec:related_work} reviews the state of the art related to emotion detection for Human-Robot Interaction (HRI) and Facial Emotion Recognition (FER) systems. Section \ref{sec:method} presents the design and implementation of the proposed emotional interface, detailing the multi-face detection, emotion recognition system, and the graphical user interface. Section \ref{sec:humanoid_robot} describes the integration of the facial emotion recognition system into the Tiago++ humanoid robot, highlighting the processes of face tracking and real-time emotion detection. Section \ref{sec:exp_results} outlines the experimental setup and presents the results, including performance metrics, model comparisons, and user interaction analysis. Finally, Section \ref{sec:conclusion} concludes the paper with a discussion of the findings, limitations of the current approach, and potential directions for future work.


\section{RELATED WORK}
\label{sec:related_work}
Although emotions have been investigated in the context of HRI, it remains a significant challenge. In this section, we report recent research in HRI as well as FER systems. 

\subsection{Emotion detection for HRI}
In social robotics, emotion detection is mimicked by robots to interact naturally and harmoniously with humans. 
Several studies have focused on implementing facial emotion recognition in robots. For instance, the study \cite{tiis:23394} applied facial emotion recognition on three datasets: FER2013, FERPLUS and FERFIN.  
The system was implemented on a NAO robot, which responds with actions based on the detected emotions. However, this study has some limitations, as it does not provide details on the robot's implementation. 
Additionally, the research \cite{dwijayanti2022real} integrated a facial detection system with a facial emotion recognition system and implemented it in a robot. They also explored automatic detection of the distance between the camera of the robot and the person. One drawback is that the robot is stationary, so mobility is not considered.
The study \cite{Spezialetti2020} serves as a survey of emotion recognition research for human-robot interaction. It reviews emotion recognition models, datasets, and modalities, with a particular emphasis on facial emotion recognition. However, it does not include any research utilizing deep learning models for facial emotion recognition. 

\subsection{Facial emotion recognition}
Deep learning has revolutionized computer vision tasks, including Facial Emotion Recognition (FER), with numerous studies proposing various methodologies to achieve high classification accuracy using well known benchmark datasets \cite{farhat2024cg,goodfellow2013challenges,letaifa2019first,mollahosseini2017affectnet,justo2021spanish,lucey2010extended}.

Several recent studies have proposed innovative approaches for FER. Farzaneh et al. \cite{farzaneh2021facial} introduced the Deep Attentive Center Loss (DACL) method, which integrates an attention mechanism to enhance feature discrimination, showing superior performance on RAF-DB and AffectNet datasets. Similarly, Pecoraro et al. \cite{pecoraro2022local} proposed the LHC-Net architecture, which employs a multi-head self-attention module tailored for FER tasks, achieving state-of-the-art results on FER2013 with lower computational complexity. In another work, Han et al. \cite{han2022triple} presented a triple-structure network model based on MobileNet V1, which captures inter-class and intra-class diversity features, demonstrating strong results on KDEF, MMI, and CK+ datasets. Fard et al. \cite{fard2022ad} introduced the Adaptive Correlation (Ad-Corre) Loss, which improved performance on AffectNet, RAF-DB, and FER2013 datasets when applied to Xception and ResNet50 models. Other notable contributions include the Segmentation VGG-19 model \cite{vignesh2023novel}, which enhanced FER on FER2013 using segmentation-inspired blocks, and the DDAMFN network by Zhang et al. \cite{zhang2023dual}, which incorporated dual-direction attention to achieve excellent results on AffectNet and FERPlus. Lastly, in our recent work, we introduced EmoNeXt \cite{el2023emonext}, a deep learning framework that has set new state-of-the-art benchmarks on the FER2013 dataset. EmoNeXt integrates a Spatial Transformer Network (STN) for handling facial alignment variations, along with Squeeze-and-Excitation (SE) blocks for channel-wise feature recalibration. Additionally, a self-attention regularization term was introduced to enhance compact feature generation, further improving accuracy. 

This  brief review shows that many FER models have focused exclusively on improving accuracy. 
As a result, today’s leading models can reach memory sizes in the order of gigabytes, which poses challenges for deployment in memory-constrained environments, such as the robots.

\section{THE EMOTIONAL INTERFACE}
\label{sec:method}
One of the  challenges in the domain of emotion detection for HRI,  is the simultaneous detection of emotions from multiple faces, which is useful  where robots interact with groups of people.

\subsection{Multi-face detection}
\label{Haar_cascade}

We choose the Haarcascade classifier, proposed by Paul Viola and Michael Jones in their seminal paper \cite{viola2001rapid}, as a highly effective method for face detection. Other notable methods include the Histogram of Oriented Gradients (HOG) combined with Support Vector Machines (SVM) and deep learning approaches such as the Multi-task Cascaded Convolutional Networks (MTCNN). While these methods have shown promising results in various applications, the Haarcascade classifier is particularly advantageous for real-time scenarios.

\begin{figure}[ht]
  \centering
    \includegraphics[width=0.475\textwidth]{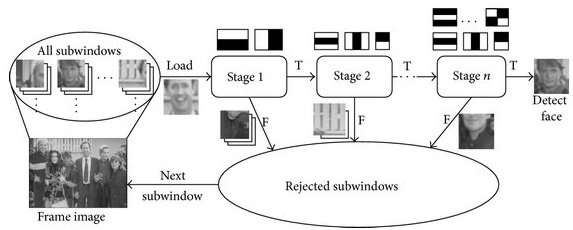}
    \caption{Cascade structure for Haar classifiers \cite{kim2015system}.}
    \label{haarcascade}
\end{figure}

The general principle of the Haarcascade approach is illustrated in Figure \ref{haarcascade}. This machine learning-based method involves training a cascade function using a large dataset of positive (face) and negative (non-face) images. The classifier relies on Haar features, which are similar to convolutional kernels, to extract distinguishing characteristics from images. Each Haar feature is a single value calculated by subtracting the sum of pixels under a white rectangle from the sum of pixels under a black rectangle. To efficiently compute these features, the concept of integral images is utilized, reducing the calculation to an operation involving just four pixels, regardless of the feature's size.

During training, all possible sizes and positions of these features are applied to the training images, resulting in over 160,000 potential features. To select the most relevant features, the AdaBoost algorithm is utilized, which iteratively adjusts the weights of misclassified images and selects features with the lowest error rates, thereby creating a strong classifier from a combination of weak classifiers. Despite the high initial number of features, this process narrows it down significantly (e.g., from 160,000 to around 6,000).

For detection, the image is scanned with a 24x24 pixel window, applying these selected features. To enhance efficiency, the authors introduced a cascade of classifiers. This means that features are grouped into stages, and if a window fails at any stage, it is immediately discarded as a non-face region. This hierarchical approach ensures that only potential face regions undergo the full, more complex evaluation process, allowing for real-time face detection with high accuracy.




\subsection{Emotion recognition system}
\label{Emotion_recognition_system}

\begin{figure}[ht!]
  \centering
    \includegraphics[width=0.55\columnwidth]{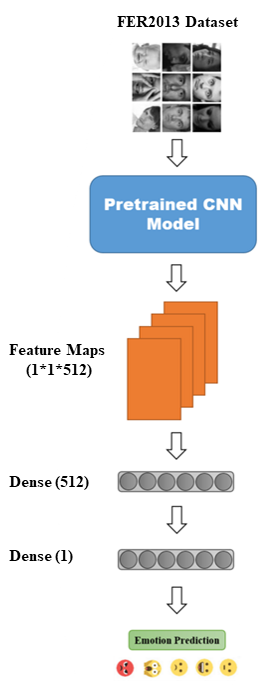}
    \caption{The architecture of the emotion recognition system using transfer learning on the FER2013 dataset.}
    \label{fig:emonext}
\end{figure}

Pretrained deep learning models have demonstrated exceptional effectiveness for feature extraction across various domains \cite{palmero2023exploring}. In our emotion recognition system (Figure. \ref{fig:emonext}), we leverage a pretrained convolutional neural network (CNN) model to apply transfer learning using the FER2013 dataset \cite{goodfellow2013challenges}.
Specifically, we utilize pretrained CNN models, initially trained on the ImageNet dataset which encompass millions of images from various categories \cite{deng2009imagenet}. This extensive training enables these models to extract highly relevant and general visual features through their convolutional layers. These layers detect fundamental elements such as edges, textures, and shapes, which are essential for understanding facial structures. We utilize these convolutional layers to process our input images, leaving out the top portion of the model, specifically the fully connected layers initially designed for the ImageNet classification tasks. Instead, by passing our facial images through the pretrained model's convolutional layers, we generate a feature stack that encapsulates essential visual information. 
This feature stack, representing a rich set of features extracted from the images, is then flattened into a format suitable for further processing. Subsequently, we introduce additional fully connected layers tailored to the FER2013 dataset to recognize and classify seven distinct emotions: anger, disgust, fear, happiness, sadness, surprise, and neutrality. 

These newly added layers are trained to fine-tune the model specifically for emotion recognition, leveraging the robust feature extraction capabilities of the pretrained model's convolutional layers.

\subsection{Graphical interface}


The graphical interface of our emotion recognition system integrates multiple advanced technologies to provide a seamless and responsive user experience. Upon launching the application, the interface is built using the Tkinter library \footnote{\url{https://docs.python.org/3/library/tkinter.html}}, creating a user-friendly graphical environment. The system activates the webcam through the OpenCV library \footnote{\url{https://docs.opencv.org/4.x/}}, capturing a live video feed for real-time analysis. Captured video frames undergo face detection using the HaarCascade classifier, a robust method for identifying faces under various lighting conditions and angles (see description in subsection \ref{Haar_cascade}).

\begin{figure}[ht!]
  \centering
    \includegraphics[width=0.32\textwidth]{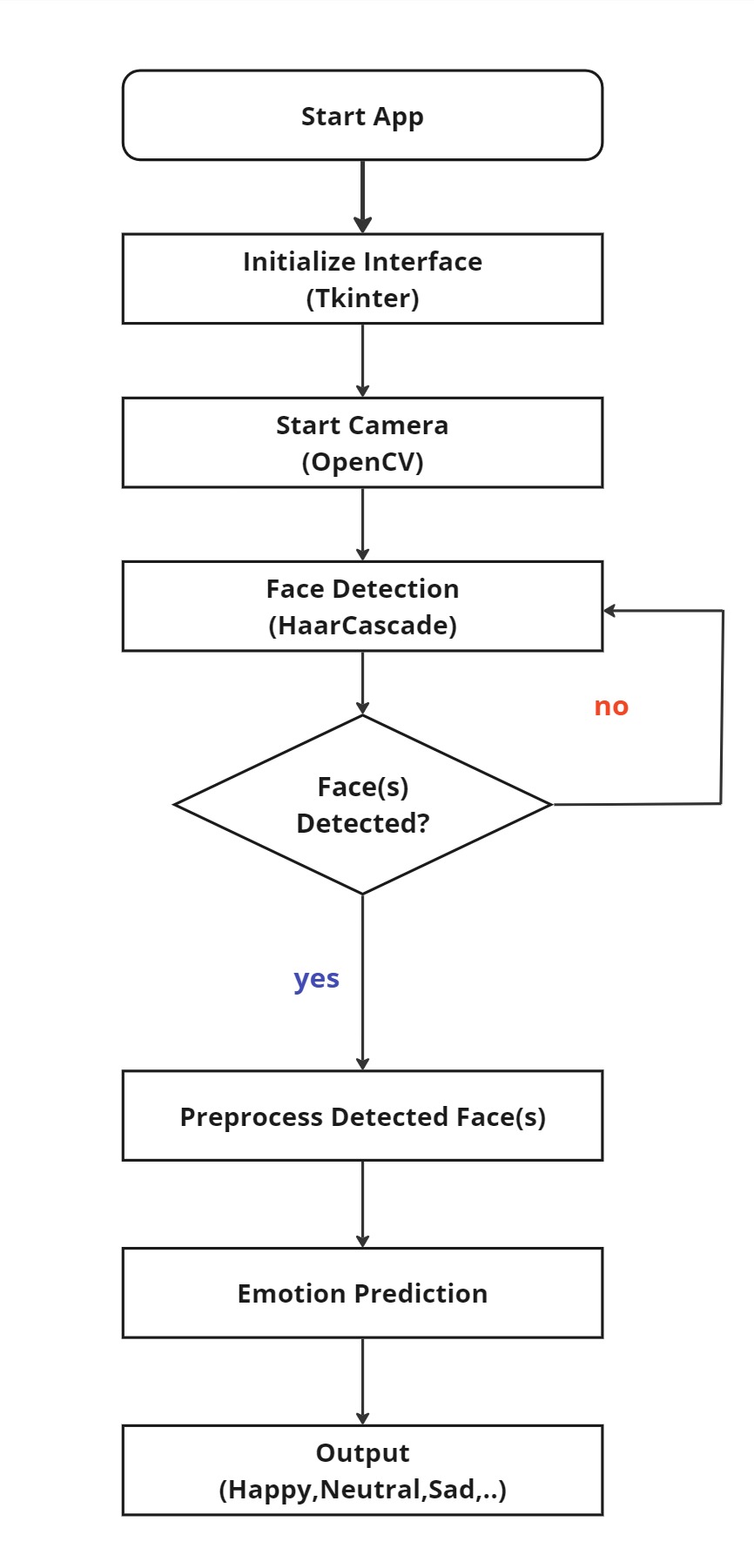}
    \caption{Global architecture of our real-time multi-face emotion recognition user interface.}
    \label{interface_architecture}
\end{figure}

Once a face is detected, the region of interest is extracted and subjected to preprocessing to ensure compatibility with the model's input size. The processed image is then fed into a pretrained CNN model that have been fine-tuned on the FER2013 dataset. 
This model analyzes the facial image to predict the user’s emotional state, categorizing it into distinct emotions such as anger, fear, disgust, happiness, sadness, surprise, and neutrality. The predicted emotion is then displayed on the graphical interface, providing immediate feedback to the user. All these steps are illustrated by Fig. \ref{interface_architecture}. 

\section{THE HUMANOID ROBOT}
\label{sec:humanoid_robot}
The Tiago robot, developed by PAL Robotics  \cite{pages2016tiago}, is a humanoid mobile robot \footnote{https://pal-robotics.com/robots/tiago/}. Its modular design allows for  customization to meet specific needs. 
In this section, we outline our approach to equipping the Tiago++ model of the robot with face emotion recognition capabilities. By using the Robot Operating System (ROS)\footnote{https://wiki.ros.org} for communication and processing, and integrating a Tkinter-based GUI for real-time visualization, we enhance the ability of the robot to interact with humans. This implementation is divided into two primary tasks: face tracking and emotion detection, each described in the following subsections.

\subsection{Face Tracking Integration on Tiago++ Robot}
\begin{figure*}[h]
    \centering
    \begin{subfigure}[b]{0.46\textwidth}
        \centering
        \includegraphics[width=1.07\linewidth,height=0.7\linewidth]{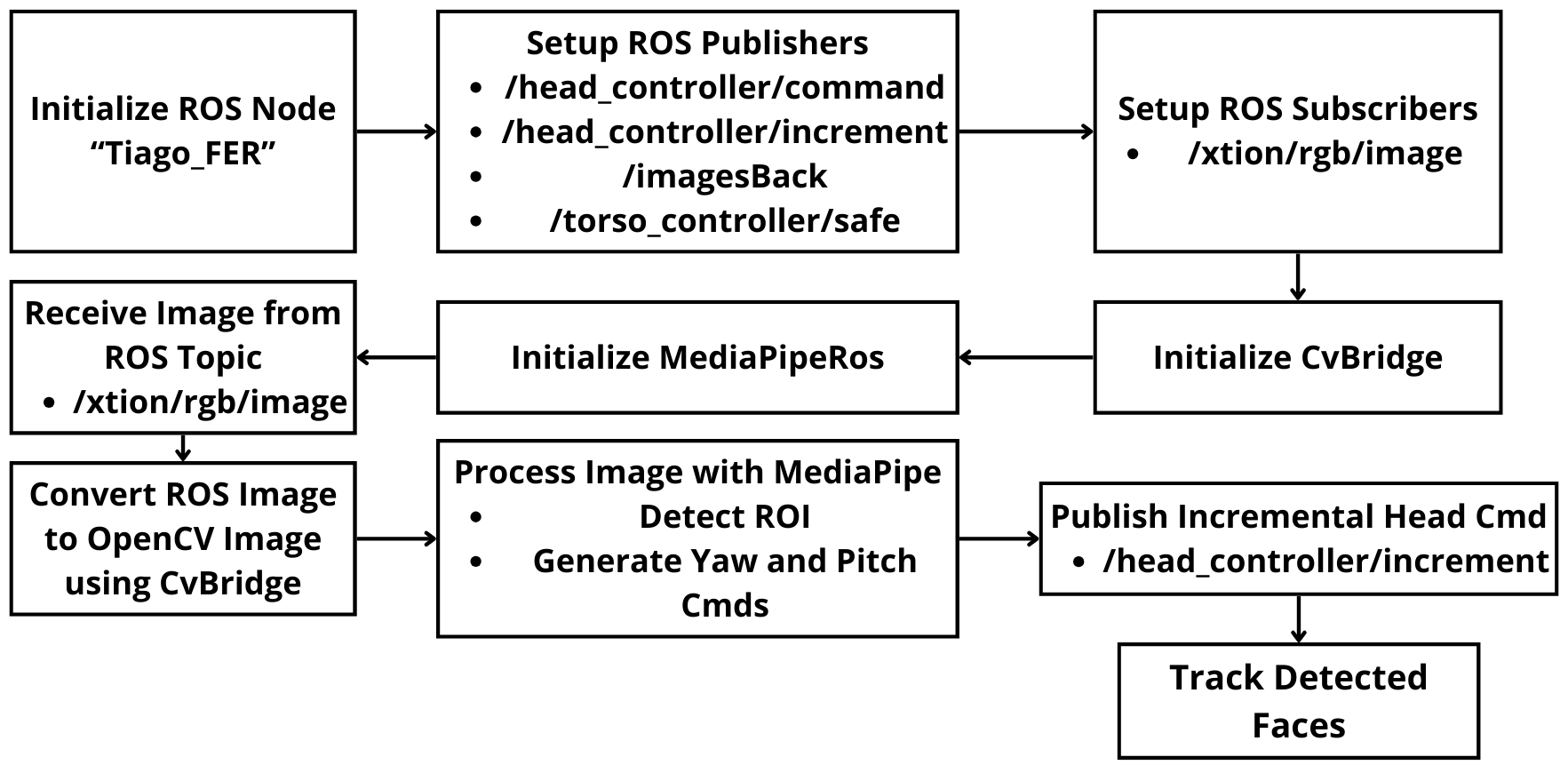}
        \caption{Face tracking integration on Tiago++ robot.}
        \label{fig:pic1}
    \end{subfigure}
    \hfill
    \begin{subfigure}[b]{0.46\textwidth}
        \centering
        \includegraphics[width=1\linewidth,height=0.5\linewidth]{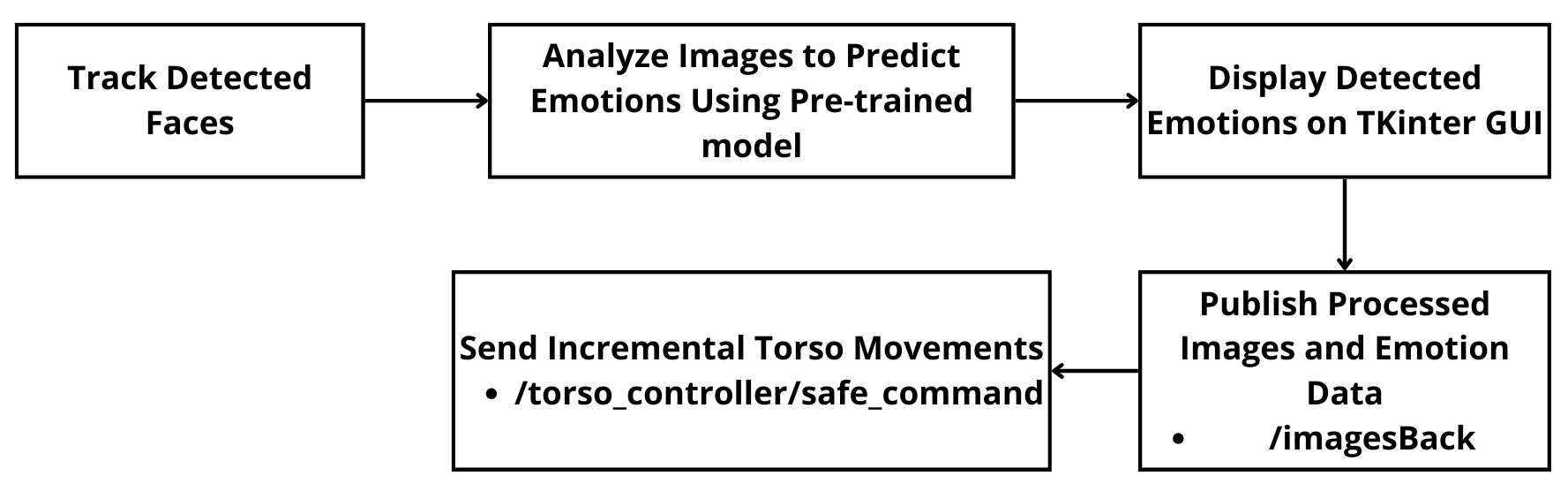}
        \caption{Emotion detection and GUI display on Tiago++ robot.}
        \label{fig:pic2}
    \end{subfigure}
    \vspace{0.5cm}
    \caption{ROS-based Tiago++ face emotion recognition integration process: the diagram in the left (a) depicts the steps involved in face tracking integration, while the diagram in the right (b) shows the emotion detection and GUI display process.}
    \label{fig:combined}
\end{figure*}

We implemented a face tracking module on the Tiago robot by integrating ROS  with a Tkinter-based GUI application. The process begins with initializing a ROS node named \texttt{Tiago\_FER} and setting up essential publishers and subscribers to facilitate communication between the robot and the software. We use the \texttt{CvBridge} \footnote{https://wiki.ros.org/cv\_bridge/Tutorials} library to convert images from ROS format to OpenCV format. Meanwhile, the \texttt{MediaPipeRos} instance processes these images to detect regions of interest (ROI) for face tracking.
The application's main loop receives images from the robot's camera through the \texttt{/xtion/rgb/image} ROS topic, processes these images to detect faces, and generates commands to adjust the robot's yaw and pitch. These commands, which control head movements, are published to the \texttt{head\_controller/increment/goal} topic using the \texttt{IncrementActionGoal} message type, enabling the robot to track the detected faces.
These steps are outlined in the diagram generated by ROS, as shown in Figure \ref{fig:combined}.

\subsection{Emotion detection and GUI display on Tiago++ screen}
Following face tracking, the processed images are analyzed to predict emotions. The detected emotions are displayed on a Tkinter GUI, which features a canvas for image display and progress bars to visualize emotion scores. The processed images and emotion data are published back to the \texttt{/imagesBack} ROS topic. 
Additionally, incremental commands for torso movements are sent to the \texttt{/torso\_controller/safe\_command} topic using the \texttt{JointTrajectory} message type, allowing the robot to dynamically respond to detected emotions (see Figure \ref{fig:combined}.

\section{EXPERIMENTS AND RESULTS}
\label{sec:exp_results}
Developing a human-robot interface for FER involves detecting faces and emotions, implementing the user interface, and integrating it into the robot platform. The robot's camera captures images of individuals interacting with it, processes these images to detect emotions, and then displays the detected emotions on the user interface. This interface is visible on the tablet mounted on the robot's chest. Several challenges are to be addressed, particularly focusing on the accuracy of the models and the feasibility of implementing them on the robot.

\subsection{Face emotion detection}
\label{sec:fer_results}
In this work, we fine-tuned several pretrained models from the Keras library\footnote{https://keras.io/api/applications/}, initially trained on the ImageNet 1000K dataset. These models were selected based on their strong performance in the ImageNet classification task and their ability to generalize well for FER tasks.
We applied transfer learning, as explained in subsection \ref{Emotion_recognition_system}, to the following models: MobileNet \cite{howard2017mobilenets}, DenseNet201 \cite{huang2017densely}, ResNet152V2 \cite{he2016identity}, ResNet101 \cite{he2016deep}, Xception \cite{chollet2017xception}, EfficientNetV2-B0 \cite{tan2021efficientnetv2}, InceptionResNetV2 and InceptionV3 \cite{szegedy2017inception}, VGG16 and VGG19 \cite{karen2014very}, and ConvNeXt (from Tiny to XLarge version) \cite{liu2022convnet}.

For training, we consistently used data augmentation techniques such as rotation, shift, zoom, horizontal flip and adjustments in brightness and contrast to improve the model's robustness. Additionally, Random Erasing was used to simulate occlusions, while resizing and recropping variations improved robustness to differences in face positioning. The models were optimized using Adam with a learning rate of 0.0001, combined with strategies like EarlyStopping and ReduceLROnPlateau to prevent overfitting and dynamically adjust the learning rate.

The accuracy and memory footprint of each fine-tuned model on the FER2013 dataset are reported in Table \ref{tab:fer_accuracy_table}. While ConvNeXt XLarge achieved the highest accuracy at 72.27\%, it comes with a significantly larger memory footprint than the other models. 

\begin{table}[htbp]
    \centering
    \caption{Pretrained models fine-tuned on the FER2013 dataset: accuracy (\%) and memory footprint (Megabytes).}
    \begin{tabular}{|l|c|c|}
    \hline
    Model name & Accuracy & Model size \\
    \hline
    \hline
        MobileNet  & 66.11 & 14.5\\
        \hline
        ResNet152V2  & 67.28 & 611.3 \\
        \hline
        DenseNet201 & 67.84 & 221.0 \\
        \hline
        InceptionV3 & 68.43 & 268.6 \\
        \hline              
        Xception  & 68.93 & 346.9 \\
        \hline        
        ConvNeXt Tiny & 69.43 & 362 \\
        \hline        
        EfficientNetV2-B0  & 70.00 & 139.0 \\
        \hline
        ConvNeXt Small & 70.15 & 566 \\
        \hline        
        InceptionResNetV2  & 70.29 & 648.2 \\
        \hline        
        ConvNeXt Base & 70.32 & 1120 \\
        \hline        
        VGG16  & 71.18 & 171.0\\
        \hline        
        ResNet101 & 71.30 & 549.8\\
        \hline  
        VGG19 & 71.46 & 262.5 \\
        \hline
        ConvNeXt Large & 71.57 & 2733 \\
        \hline        
        \textbf{ConvNeXt XLarge} & \textbf{72.27} & \textbf{3900} \\
    \hline
    \end{tabular}
    \label{tab:fer_accuracy_table}
\end{table}

\subsection{Confidence interval}
Accuracy is an estimate of the performance of a system, and its reliability depends on the number of tests conducted that is in our case the number of emotions to be recognized. The measurement of the confidence interval is introduced to assess the trustability of our recognition rate. In \cite{zouari2007vers}, the successes are modeled by a binomial distribution. If N is the number of tests  and P is the recognition rate, then the confidence interval [P-, P+] at x\% is:

\begin{figure}[h]
  \centering
    \includegraphics[width=0.38\textwidth]{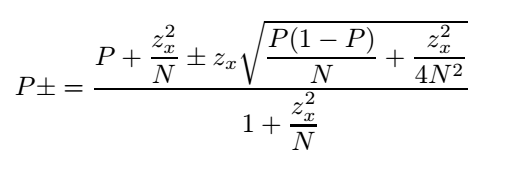}
    \label{Confidence_interval}
\end{figure}

with z95\% = 1.96 and z98\% = 2.33. This means that there is a x\% chance that the rate falls within the interval [P-, P+].

\begin{figure}[ht]
  \centering
    \includegraphics[angle=270,origin=c, width=0.48\textwidth]{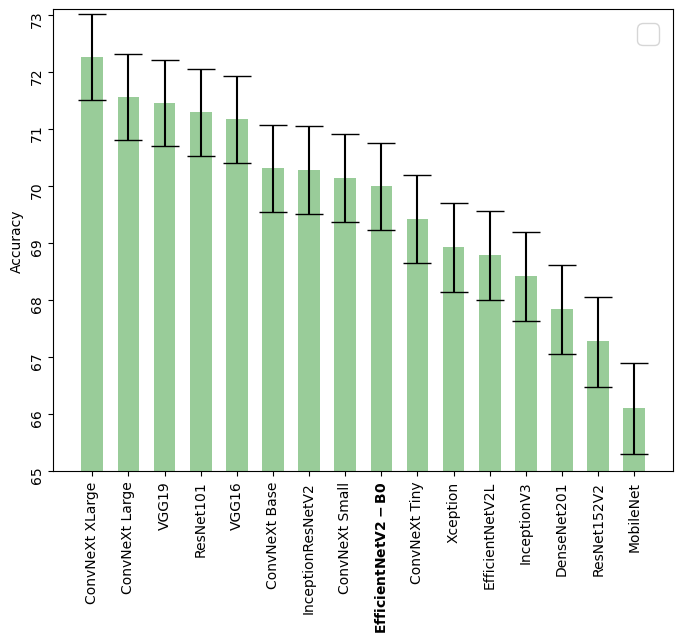}
    \caption{Accuracy and confidence intervals of the  models}
    \label{error_bars}
\end{figure}

The FER2013 dataset consists in 35,887 grayscale
images, divided into training (80\%), test (10\%) and validation (10\%). Hence, using each model 3589 samples have been evaluated on the test set. We compute the confidence interval with z98 for all models and report the results in Figure \ref{error_bars}. 
We notice that several models, including VGG16, InceptionResNetV2, ConvNeXt Base, EfficientNetV2-B0, and VGG19, show overlapping results. In terms of precision, these models demonstrate similar efficiency. However, there is a notable difference in their sizes, with EfficientNetV2-B0 being the most compact. Due to its smaller size, EfficientNetV2-B0 has been chosen for implementation on the robot.

\subsection{User interface development}
Our emotion recognition application features an intuitive and user-friendly graphical interface designed for both single-face and multi-face emotion detection. The interface allows users to utilize their device's camera to capture live video streams, which are then processed in real-time to detect and classify facial expressions. For single-face emotion recognition, the application highlights the detected face and displays the identified emotion with corresponding confidence levels. In multi-face scenarios, the interface efficiently detects multiple faces within the same frame, assigning emotions to each detected face individually. The results are visually presented using bounding boxes and emotion labels directly on the video feed, providing clear and immediate feedback.

\begin{figure}[ht]
  \centering
    \includegraphics[width=0.4\textwidth]{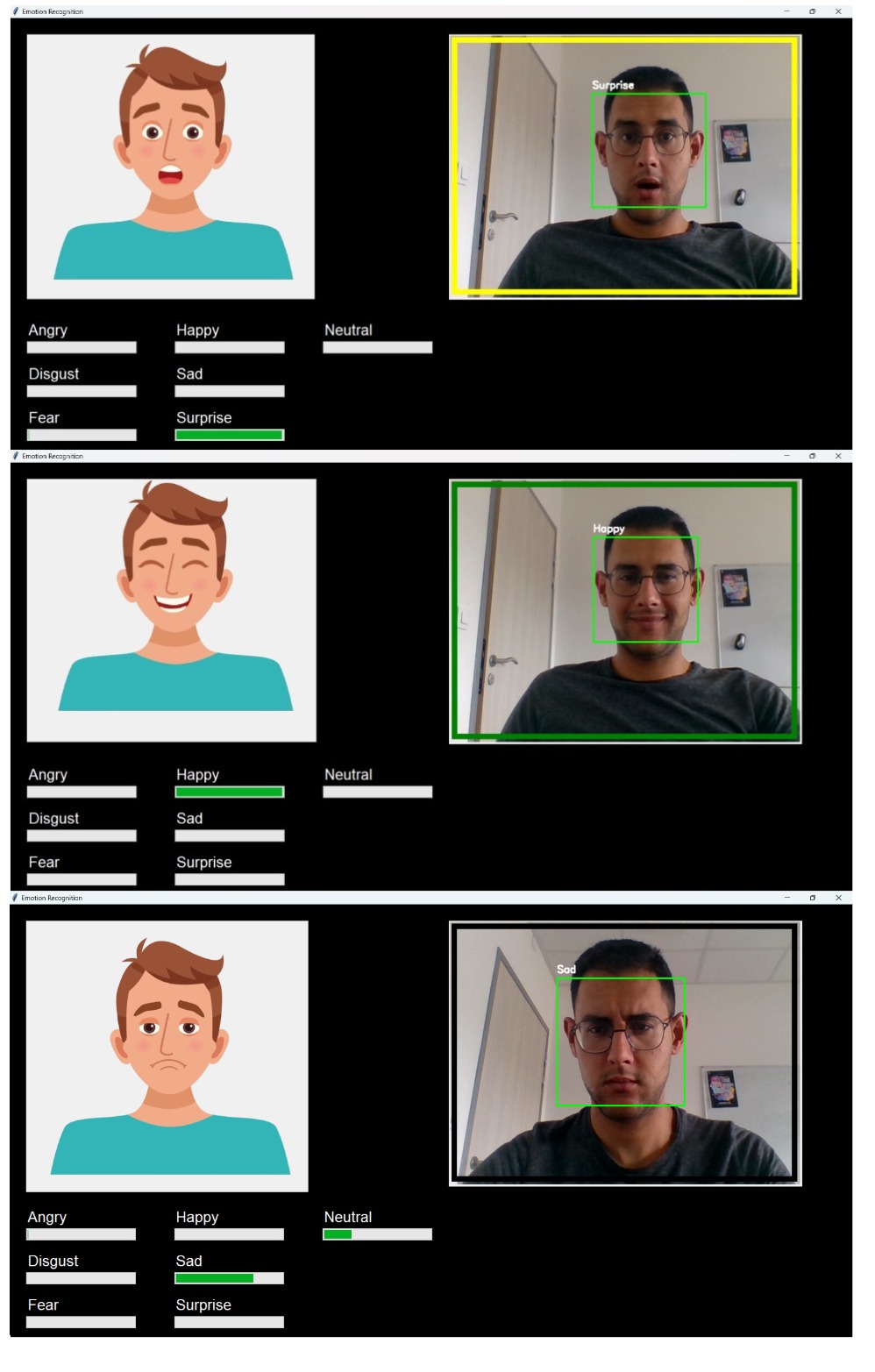}
    \caption{The user interface displays face and emotion detection for a single person. Progress bars indicate the confidence score for each recognized emotion.}
    \label{face_detection}
\end{figure}

\begin{figure}[ht]
  \centering
    \includegraphics[width=0.3\textwidth]{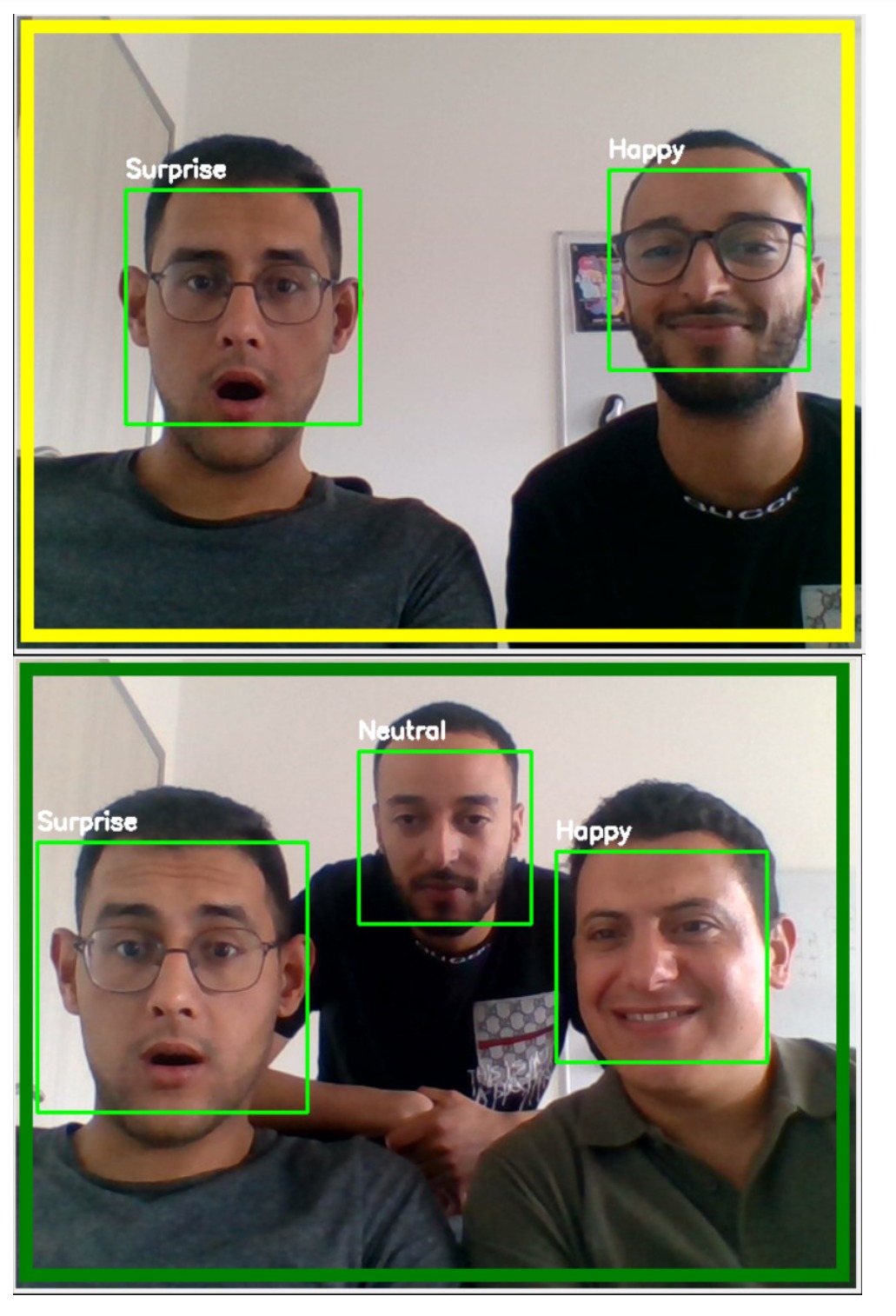}
    \caption{Face detection is followed by emotion detection for multiple individuals present in the same image.}
    \label{multiface_detection}
\end{figure}

Additionally, the interface includes progress bars for the detected emotion, visually representing the confidence level of each prediction. An avatar further enhances user interaction by imitating the predicted emotion in real-time, offering an engaging and dynamic way to understand the results. This comprehensive and interactive interface ensures that users can easily interpret the emotion detection outcomes, making the application practical for various real-world settings, including human-robot interaction and affective computing.
Figure. \ref{face_detection} and Figure. \ref{multiface_detection} show some examples of the user interface applied to single and multi-face emotion detection.

\subsection{FER deployment on Tiago++ robot}

The Tiago++ is a humanoid mobile robot with constrained resources (CPU, memory, and storage). Besides interacting with humans, the robot must concurrently perform critical tasks such as navigation and detection, which are also resource-intensive.
Consequently, for deploying our application on the Tiago++ robot, it is essential to select a model not only based on its test accuracy but also on the memory footprint of the model. The Tiago++ robot has a maximum capacity of about 150 MB for model files to ensure real-time inference without disrupting other processes running on the robot. According to  Table \ref{tab:fer_accuracy_table} and the previous subsection, EfficientNetV2-B0 stands out with a good balance between accuracy (70.00\%) and model size (139 MB), meeting the robot's constraints.

\begin{figure}
  \centering
    \includegraphics[width=0.35\textwidth]{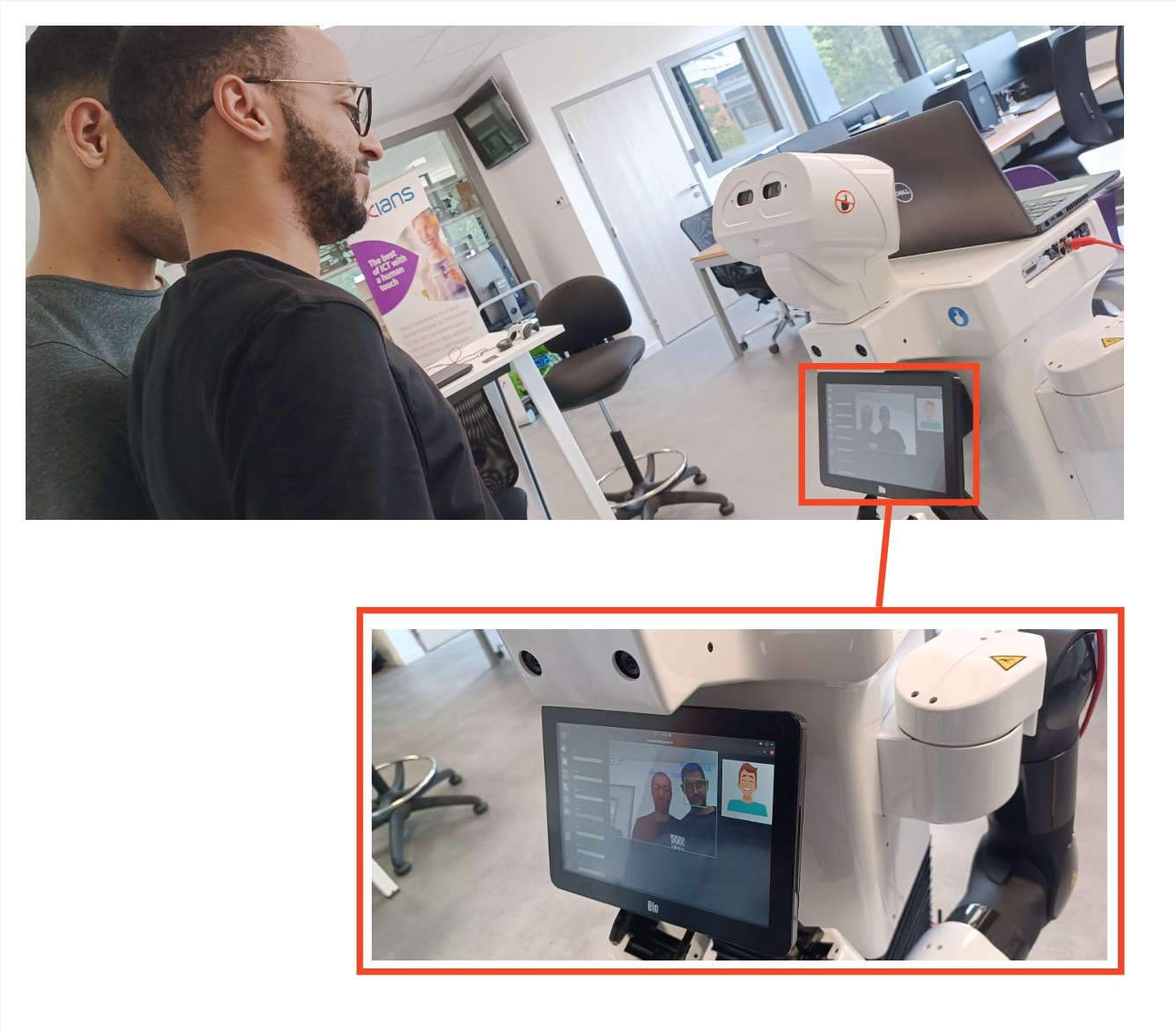}
    \caption{Multi face emotion  detection deployed on robot}
    \label{robot_interface}
\end{figure}

To illustrate the system's effectiveness, we conducted two sets of experiments. In the first set, a single participant interacted with the robot, displaying a range of emotions. The system's ability to accurately detect the face and classify the emotional state of the participant in real-time was meticulously observed and documented. In the second set, two participants were present simultaneously, engaging in various interactions with the robot. This scenario tested the system's robustness in detecting multiple faces and correctly identifying each individual's emotional state in real-time.
The results of these experiments are depicted through a series of images captured during the interactions on Figure. \ref{robot_interface}. 

\section{CONCLUSIONS} 
\label{sec:conclusion}
In this paper, we presented a facial emotion detection interface implemented on a mobile humanoid robot. This interface is capable of displaying emotions from multiple individuals in real-time video. To achieve this, we developed and evaluated several deep neural network models under consistent conditions, carefully considering factors such as model size and accuracy to ensure compatibility with both personal computers and mobile robots like the Tiago++.

While our system demonstrates strong performance, it is important to note the limitations of relying solely on facial expressions for emotion detection, particularly in contexts where communication may be impaired. Emotions are complex and multifaceted, often requiring the integration of multiple modalities for more accurate recognition. Therefore, future work will focus on incorporating additional modalities, such as voice, text, gestures, and biosignals, to enhance the performance and reliability of emotion recognition systems. Additionally, we will focus on optimizing large models used in FER tasks to ensure their efficiency for deployment on the Tiago++ robot, considering the balance between model size and accuracy.

\bibliographystyle{apalike}
{\small
\bibliography{biblio}}

\begin{thebibliography}{}

\bibitem[Chollet, 2017]{chollet2017xception}
Chollet, F. (2017).
\newblock Xception: Deep learning with depthwise separable convolutions.
\newblock In {\em Proceedings of the IEEE conference on computer vision and pattern recognition}, pages 1251--1258.

\bibitem[Deng et~al., 2009]{deng2009imagenet}
Deng, J., Dong, W., Socher, R., Li, L.-J., Li, K., and Fei-Fei, L. (2009).
\newblock Imagenet: A large-scale hierarchical image database.
\newblock In {\em 2009 IEEE conference on computer vision and pattern recognition}, pages 248--255. Ieee.

\bibitem[Dwijayanti et~al., 2022]{dwijayanti2022real}
Dwijayanti, S., Iqbal, M., and Suprapto, B.~Y. (2022).
\newblock Real-time implementation of face recognition and emotion recognition in a humanoid robot using a convolutional neural network.
\newblock {\em IEEE Access}, 10:89876--89886.

\bibitem[El~Boudouri and Bohi, 2023]{el2023emonext}
El~Boudouri, Y. and Bohi, A. (2023).
\newblock Emonext: an adapted convnext for facial emotion recognition.
\newblock In {\em 2023 IEEE 25th International Workshop on Multimedia Signal Processing (MMSP)}, pages 1--6. IEEE.

\bibitem[Fard and Mahoor, 2022]{fard2022ad}
Fard, A.~P. and Mahoor, M.~H. (2022).
\newblock Ad-corre: Adaptive correlation-based loss for facial expression recognition in the wild.
\newblock {\em IEEE Access}, 10:26756--26768.

\bibitem[Farhat et~al., 2024]{farhat2024cg}
Farhat, N., Bohi, A., Letaifa, L.~B., and Slama, R. (2024).
\newblock Cg-mer: a card game-based multimodal dataset for emotion recognition.
\newblock In {\em Sixteenth International Conference on Machine Vision (ICMV 2023)}, volume 13072, pages 399--406. SPIE.

\bibitem[Farzaneh and Qi, 2021]{farzaneh2021facial}
Farzaneh, A.~H. and Qi, X. (2021).
\newblock Facial expression recognition in the wild via deep attentive center loss.
\newblock In {\em Proceedings of the IEEE/CVF winter conference on applications of computer vision}, pages 2402--2411.

\bibitem[Goodfellow et~al., 2013]{goodfellow2013challenges}
Goodfellow, I.~J., Erhan, D., Carrier, P.~L., Courville, A., Mirza, M., Hamner, B., Cukierski, W., Tang, Y., Thaler, D., Lee, D.-H., et~al. (2013).
\newblock Challenges in representation learning: A report on three machine learning contests.
\newblock In {\em Neural information processing. 20th international conference, ICONIP}, pages 117--124. Springer.

\bibitem[Gouaillier et~al., 2009]{Gouaillier_2009}
Gouaillier, D., Hugel, V., Blazevic, P., and Kilner, C. (2009).
\newblock Mechatronic design of nao humanoid.
\newblock In {\em IEEE International Conference on Robotics and Automation ICRA}.

\bibitem[Han et~al., 2022]{han2022triple}
Han, B., Hu, M., Wang, X., and Ren, F. (2022).
\newblock A triple-structure network model based upon mobilenet v1 and multi-loss function for facial expression recognition.
\newblock {\em Symmetry}, 14(10):2055.

\bibitem[He et~al., 2016a]{he2016deep}
He, K., Zhang, X., Ren, S., and Sun, J. (2016a).
\newblock Deep residual learning for image recognition.
\newblock In {\em Proceedings of the IEEE conference on computer vision and pattern recognition}, pages 770--778.

\bibitem[He et~al., 2016b]{he2016identity}
He, K., Zhang, X., Ren, S., and Sun, J. (2016b).
\newblock Identity mappings in deep residual networks.
\newblock In {\em Computer Vision--ECCV 2016: 14th European Conference, Amsterdam, The Netherlands, October 11--14, 2016, Proceedings, Part IV 14}, pages 630--645. Springer.

\bibitem[Hirose and Ogawa, 2007]{hirose_2007}
Hirose, M. and Ogawa, K. (2007).
\newblock Honda humanoid robots development.
\newblock {\em Philosophical Transactions of the Royal Society A: Mathematical, Physical and Engineering Sciences}, 365(1850):11--19.

\bibitem[Howard et~al., 2017]{howard2017mobilenets}
Howard, A.~G., Zhu, M., Chen, B., Kalenichenko, D., Wang, W., Weyand, T., Andreetto, M., and Adam, H. (2017).
\newblock Mobilenets: Efficient convolutional neural networks for mobile vision applications.
\newblock {\em arXiv preprint arXiv:1704.04861}.

\bibitem[Huang et~al., 2017]{huang2017densely}
Huang, G., Liu, Z., Van Der~Maaten, L., and Weinberger, K.~Q. (2017).
\newblock Densely connected convolutional networks.
\newblock In {\em Proceedings of the IEEE conference on computer vision and pattern recognition}, pages 4700--4708.

\bibitem[Justo et~al., 2021]{justo2021spanish}
Justo, R., Letaifa, L.~B., Olaso, J.~M., L{\'o}pez-Zorrilla, A., Develasco, M., V{\'a}zquez, A., and Torres, M.~I. (2021).
\newblock A spanish corpus for talking to the elderly.
\newblock {\em Conversational Dialogue Systems for the Next Decade}, pages 183--192.

\bibitem[Justo et~al., 2020]{Justo2020}
Justo, R., Letaifa, L.~B., Palmero, C., Fraile, E.~G., Johansen, A., Vazquez, A., Cordasco, G., Schlogl, S., Ruanova, B.~F., Silva, M., Escalera, S., Velasco, M.~D., Laranga, J.~T., Esposito, A., Kornes, M., and Torres, M.~I. (2020).
\newblock Analysis of the interaction between elderly people and a simulated virtual coach.
\newblock {\em Journal of Ambient Intelligence and Humanized Computing}, 11:6125--6140.

\bibitem[Karen, 2014]{karen2014very}
Karen, S. (2014).
\newblock Very deep convolutional networks for large-scale image recognition.
\newblock {\em arXiv preprint arXiv: 1409.1556}.

\bibitem[Kim et~al., 2015]{kim2015system}
Kim, M., Lee, D., and Kim, K.-Y. (2015).
\newblock System architecture for real-time face detection on analog video camera.
\newblock {\em International Journal of Distributed Sensor Networks}, 11(5):251386.

\bibitem[Letaifa et~al., 2019]{letaifa2019first}
Letaifa, L.~B., Develasco, M., Justo, R., and Torres, M.~I. (2019).
\newblock First steps to develop a corpus of interactions between elderly and virtual agents in spanish with emotion.
\newblock In {\em International Conference on Statistical Language and Speech Processing}.

\bibitem[Liu et~al., 2022]{liu2022convnet}
Liu, Z., Mao, H., Wu, C.-Y., Feichtenhofer, C., Darrell, T., and Xie, S. (2022).
\newblock A convnet for the 2020s.
\newblock In {\em Proceedings of the IEEE/CVF conference on computer vision and pattern recognition}, pages 11976--11986.

\bibitem[Lucey et~al., 2010]{lucey2010extended}
Lucey, P., Cohn, J.~F., Kanade, T., Saragih, J., Ambadar, Z., and Matthews, I. (2010).
\newblock The extended cohn-kanade dataset (ck+): A complete dataset for action unit and emotion-specified expression.
\newblock In {\em 2010 ieee computer society conference on computer vision and pattern recognition-workshops}, pages 94--101. IEEE.

\bibitem[Mollahosseini et~al., 2017]{mollahosseini2017affectnet}
Mollahosseini, A., Hasani, B., and Mahoor, M.~H. (2017).
\newblock Affectnet: A database for facial expression, valence, and arousal computing in the wild.
\newblock {\em IEEE Transactions on Affective Computing}, 10(1):18--31.

\bibitem[Olaso et~al., 2021]{olaso2021}
Olaso, J., V{\'a}zquez, A., Letaifa, L.~B., de~Velasco, M., Mtibaa, A., Hmani, M.~A., Petrovska-Delacr{\'e}taz, D., Chollet, G., Montenegro, C., L{\'o}pez-Zorrilla, A., et~al. (2021).
\newblock The empathic virtual coach: a demo.
\newblock In {\em The 2021 International Conference on Multimodal Interaction (ICMI'21)}, pages 848--851. ACM.

\bibitem[Pages et~al., 2016]{pages2016tiago}
Pages, J., Marchionni, L., and Ferro, F. (2016).
\newblock Tiago: the modular robot that adapts to different research needs.
\newblock In {\em International workshop on robot modularity, IROS}, volume 290.

\bibitem[Palmero et~al., 2023]{palmero2023exploring}
Palmero, C., deVelasco, M., Amine~Hmani, M., Mtibaa, A., Ben~Letaifa, L., Buch-Cardona, P., Justo, R., Amorese, T., Gonz{\'a}lez-Fraile, E., Fern{\'a}ndez-Ruanova, B., et~al. (2023).
\newblock Exploring emotion expression recognition in older adults interacting with a virtual coach.
\newblock {\em arXiv e-prints}, pages arXiv--2311.

\bibitem[Pecoraro et~al., 2022]{pecoraro2022local}
Pecoraro, R., Basile, V., and Bono, V. (2022).
\newblock Local multi-head channel self-attention for facial expression recognition.
\newblock {\em Information}, 13(9):419.

\bibitem[Peltier and Fiorini, 2017]{peltier_2017}
Peltier, A. and Fiorini, L. (2017).
\newblock Buddy: A companion robot for living assistance.
\newblock {\em Journal of Robotics and Automation}, 3(2):75--81.

\bibitem[Spezialetti et~al., 2020]{Spezialetti2020}
Spezialetti, M., Placidi, G., and Rossi, S. (2020).
\newblock Emotion recognition for human-robot interaction: Recent advances and future perspectives.
\newblock {\em Frontiers in Robotics and AI}, 7:145.

\bibitem[Szegedy et~al., 2017]{szegedy2017inception}
Szegedy, C., Ioffe, S., Vanhoucke, V., and Alemi, A. (2017).
\newblock Inception-v4, inception-resnet and the impact of residual connections on learning.
\newblock In {\em Proceedings of the AAAI conference on artificial intelligence}, volume~31.

\bibitem[Tan and Le, 2021]{tan2021efficientnetv2}
Tan, M. and Le, Q. (2021).
\newblock Efficientnetv2: Smaller models and faster training.
\newblock In {\em International conference on machine learning}, pages 10096--10106. PMLR.

\bibitem[Vignesh et~al., 2023]{vignesh2023novel}
Vignesh, S., Savithadevi, M., Sridevi, M., and Sridhar, R. (2023).
\newblock A novel facial emotion recognition model using segmentation vgg-19 architecture.
\newblock {\em International Journal of Information Technology}, pages 1--11.

\bibitem[Viola and Jones, 2001]{viola2001rapid}
Viola, P. and Jones, M. (2001).
\newblock Rapid object detection using a boosted cascade of simple features.
\newblock In {\em Proceedings of the IEEE computer society conference on computer vision and pattern recognition. CVPR}, volume~1, pages I--I. Ieee.

\bibitem[Zhang et~al., 2023]{zhang2023dual}
Zhang, S., Zhang, Y., Zhang, Y., Wang, Y., and Song, Z. (2023).
\newblock A dual-direction attention mixed feature network for facial expression recognition.
\newblock {\em Electronics}, 12(17):3595.

\bibitem[Zhao et~al., 2020]{tiis:23394}
Zhao, G., Yang, H., Tao, Y., Zhang, L., and and, C.~Z. (2020).
\newblock Lightweight cnn-based expression recognition on humanoid robot.
\newblock {\em KSII Transactions on Internet and Information Systems}, 14(3):1188--1203.

\bibitem[Zouari, 2007]{zouari2007vers}
Zouari, L. (2007).
\newblock {\em Vers le temps r{\'e}el en transcription automatique de la parole grand vocabulaire}.
\newblock PhD thesis, T{\'e}l{\'e}com ParisTech.

\end{thebibliography}

\end{document}